\documentclass[aps,prl,
twocolumn,
showpacs,amsmath,amssymb,
superscriptaddress,longbibliography]{revtex4-2}
\usepackage{amsmath}
\usepackage{amsfonts}
\usepackage{amssymb}
\usepackage{graphicx}
\graphicspath{{./figures/}}
\usepackage{epsfig}
\usepackage{color}
\usepackage{empheq}
\usepackage{braket}
\usepackage{epstopdf}
\definecolor{gold}{RGB}{255,215,0}
\definecolor{blue}{RGB}{0,0,255}

\begin{document}
\title{Nanomechanical damping via electron-assisted relaxation of two-level systems}

\author{Olivier Maillet}
\email[]{olivier.maillet@cea.fr}
\affiliation{Universit\'e Grenoble Alpes, CNRS, Institut N\'eel, 38000 Grenoble, France}
\affiliation{Universit\'e Paris-Saclay, CEA, CNRS, SPEC, 91191 Gif-sur-Yvette Cedex, France}
\author{Dylan Cattiaux}
\affiliation{Universit\'e Grenoble Alpes, CNRS, Institut N\'eel, 38000 Grenoble, France}
\author{Xin Zhou}
\affiliation{Universit\'e Grenoble Alpes, CNRS, Institut N\'eel, 38000 Grenoble, France}
\affiliation{CNRS, Univ. Lille, Centrale Lille, Univ. Polytechnique Hauts-de-France,
IEMN UMR8520, Av. Henri Poincar\'e, Villeneuve d'Ascq 59650, France}
\author{Rasul R. Gazizulin}
\affiliation{Universit\'e Grenoble Alpes, CNRS, Institut N\'eel, 38000 Grenoble, France}
\author{Olivier Bourgeois}
\affiliation{Universit\'e Grenoble Alpes, CNRS, Institut N\'eel, 38000 Grenoble, France}
\author{Andrew D. Fefferman}
\affiliation{Universit\'e Grenoble Alpes, CNRS, Institut N\'eel, 38000 Grenoble, France}
\author{Eddy Collin}
\affiliation{Universit\'e Grenoble Alpes, CNRS, Institut N\'eel, 38000 Grenoble, France}

\begin{abstract}
We report on measurements of dissipation and frequency noise at millikelvin temperatures of nanomechanical devices covered with aluminum. A clear excess damping is observed after switching the metallic layer from superconducting to the normal state with a magnetic field. Beyond the standard model of internal tunneling systems coupled to the phonon bath, here we consider the relaxation to the conduction electrons together with the nature of the mechanical dispersion laws for stressed/unstressed devices. With these key ingredients, a model describing the relaxation of two-level systems inside the structure due to interactions with electrons and phonons with well separated timescales captures the data. In addition, we measure an excess $1/f$-type frequency noise in the normal state, which further emphasizes the impact of conduction electrons. 
\end{abstract}
\maketitle
\section{Introduction}
Nano-electro-mechanical systems (NEMS) \cite{Craighead2000} are now common tools used for ultra-sensitive detection \cite{chaste_nanomechanical_2012} while being ubiquitous model systems for the study of quantum foundations involving mechanical degrees of freedom \cite{oconnell_quantum_2010,Ockeloen-Korppi2018}. Both endeavours require resonators with high quality factors $Q$ \cite{Ghadimi2018}, so as to resolve either small frequency changes due to e.g. masses added \cite{chaste_nanomechanical_2012}, or to preserve quantum coherence over long enough times \cite{wilson_measurement-based_2015}. Yet, mechanisms limiting the intrinsic $Q$ factor of nanomechanical systems mostly remain a puzzle despite intensive efforts \cite{zolfagharkhani_quantum_2005,unterreithmeier_damping_2010,hoehne_damping_2010}, especially at low temperature where quantum effects are expected to manifest themselves. Commonly proposed mechanisms include clamping losses \cite{wilson-rae_intrinsic_2008}, and higher order phonon processes \cite{cleland_foundations_2002}, e.g. thermoelastic damping \cite{lifshitz_thermoelastic_2000} and Akhiezer damping \cite{Iyer2016}. While clamping losses are vanishingly small for thin beam structures \cite{wilson-rae_intrinsic_2008}, phonon-phonon interactions are switched off at low temperatures. In most cases, the surviving mechanism is thought to be the coupling between the mechanical strain arising from the resonator's motion and low-energy excitations in the constitutive material \cite{zolfagharkhani_quantum_2005,hoehne_damping_2010,venkatesan_dissipation_2010,Hauer2018}. The latter are either defects or (groups of) atoms that tunnel quantum mechanically between two nearly equivalent positions in the atomic lattice, hence forming two-level systems (TLS). These TLS cause damping of the mechanical motion through their interaction with the induced strain field and their own energy relaxation. The initial microscopic description of such a mechanism, the so-called Standard Tunneling Model (STM), was introduced in the early 70s \cite{anderson_anomalous_1972,phillips_tunneling_1972} to explain low-temperature properties of amorphous materials and is still widely used nowadays, its importance being renewed by e.g. superconducting circuits \cite{martinis_decoherence_2005,capelle2018,Muller_2019,sueur2018microscopic} or nanomechanics studies \cite{hoehne_damping_2010,venkatesan_dissipation_2010,hamoumi_microscopic_2018,Hauer2018}. However, in the latter case, the model was unsuccessful in describing nano-systems which integrate resistive metallic elements \cite{zolfagharkhani_quantum_2005,venkatesan_dissipation_2010, lulla_evidence_2013}.% In parallel, the macroscopic description of damping in nanoresonators has been clarified \cite{unterreithmeier_damping_2010}, in particular by considering the role of stress \cite{southworth_stress_2009}. 
\begin{figure}[h]
\centering
\includegraphics[width=\columnwidth]
{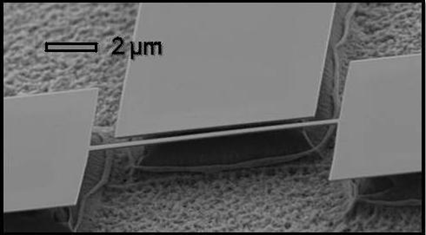}
\includegraphics[width=\columnwidth]
{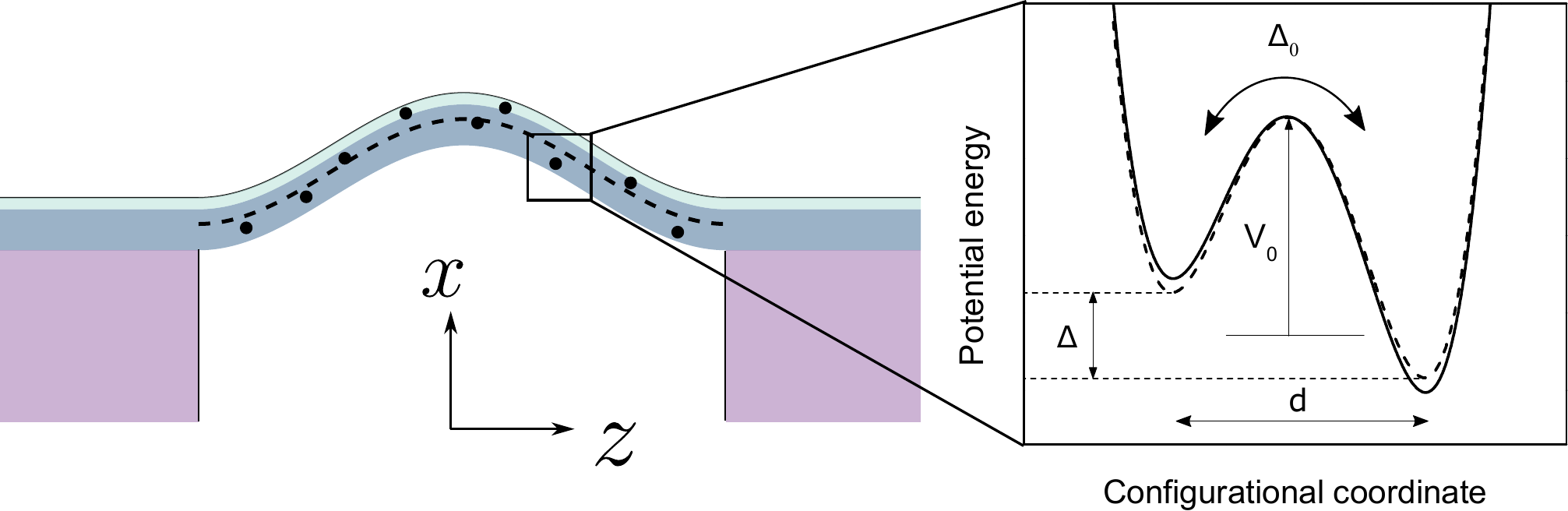}
\caption{Top: scanning electron micrograph of the device (side gate electrode not used). %The nanobeam is released through XeF$_2$ etching of the silicon lying underneath, which results in an undercut and a spongy bottom. 
Bottom: schematic side view of NEMS (left) and TLS-strain coupling (right) under a macroscopic applied force. Right panel: schematic potential energy of a single TLS at equilibrium (dashed) and under strain (full line). The two minima (separated by an energy gap $\Delta$) are coupled through a tunneling element $\Delta_0\propto e^{-d\sqrt{2m_aV_0}/\hbar}$, where $V_0$ is the barrier height, $d$ the interwell distance and $m_a$ the tunneling entity effective mass.}
\label{Fig1}
\end{figure}

In this Article, we report the measurements from 10\,\,K down to 30 mK of the damping rate and frequency shift of a high $Q$, high stress silicon nitride NEMS beam, covered with a thin aluminum layer (Fig. \ref{Fig1}). Using different magnetic fields we can tune the metallic layer state from superconducting to normal below 1 K, revealing the unambiguous contribution to nanomechanical damping of the normal state electrons reported for low-stress nanocantilevers \cite{lulla_evidence_2013}. To explain this contribution, we quantitatively include a mechanism of TLS relaxation due to the conduction electron bath \cite{hunklinger_chapter_1986} in parallel with phonon-assisted relaxation, for a given fraction of the TLS distribution. The reasoning is formally equivalent to the one proposed in \cite{unterreithmeier_damping_2010} where the authors introduce ad hoc a retarded (imaginary in frequency domain) Young's modulus, whose microscopic origin is addressed in our work. The data are fit in all regimes, with a minimal set of free parameters. For comparison, the model also successfully reproduces the nanocantilever data of Ref. \cite{lulla_evidence_2013}. Details on the calculations and additional data can be found in the Supplementary material (see \cite{Suppmat} and references \cite{pohl2002,regal08,zhou2019} therein). In addition, we measure the frequency noise as a function of temperature in both normal and superconducting state. The magnitude is found to differ substantially between the two states, pointing again towards an electron-assisted mechanism. The results together with the model provide an answer to the issue of nano-electro-mechanical damping in (hybrid) metallic systems at low temperatures, consistent with all the related results reported in the literature so far.
\section{Experimental results}
\begin{figure*}
	\centering
	\includegraphics[width=17.2cm]{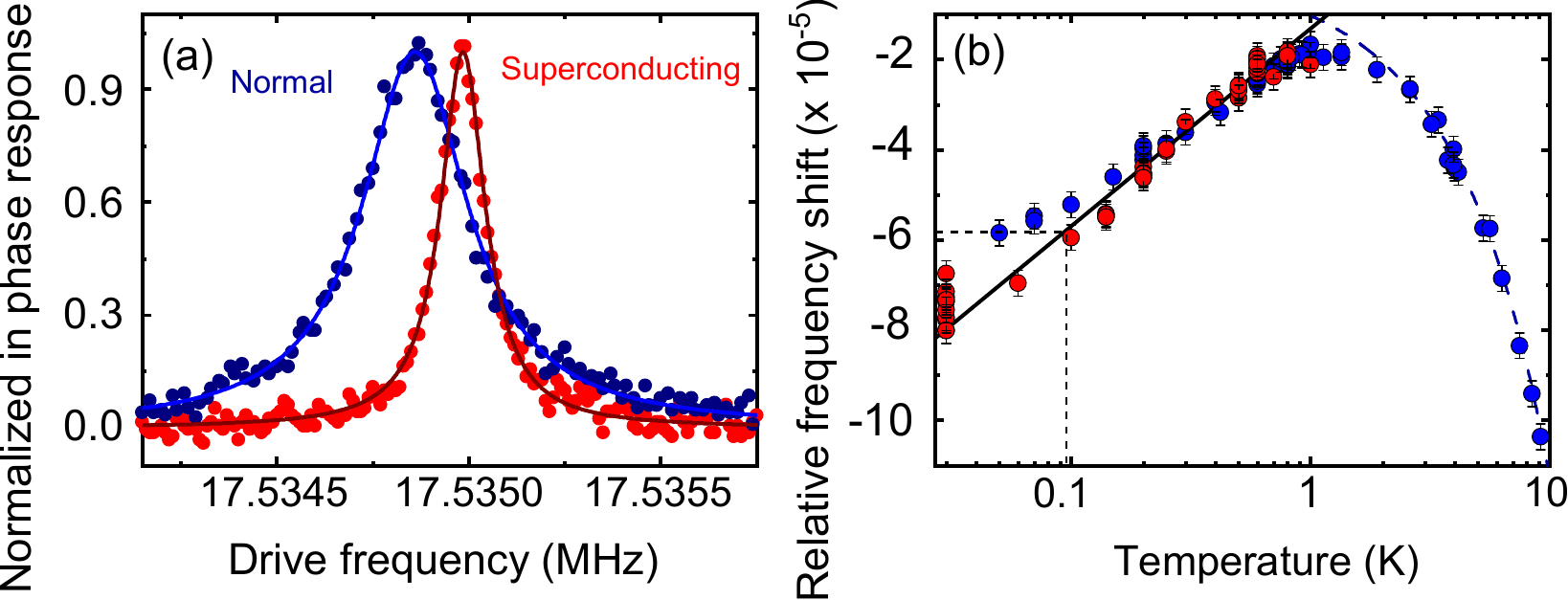}
	\caption{a) NEMS in-phase response to a small magnetomotive excitation at $T=275$ mK in superconducting (red dots) and normal (blue dots) states of the metallic layer, normalized to their peak height for better comparison. Solid lines are Lorentzian fits. The resonance frequency difference is due to the magnetomotive contribution, and the magnetomotive damping contribution accounts for $5$ \% of the line width in the normal state. b) Relative frequency shift $\delta\omega_0/\omega_0$ as a function of refrigerator's base temperature. The line is a logarithmic fit, while the dashed blue curve is a $T^3$ empirical law (see text). The dotted line indicates that, by using the frequency shift as a thermometer, the electron temperature could be about 100 mK when the cryostat's base temperature is 50 mK (see text).}
	\label{Fig2}
\end{figure*}

The main sample is a 15 $\mu$m long silicon nitride beam covered with $30$ nm aluminum, having transverse dimensions $e\times w =$ 130 nm $\times$ 300 nm (see Fig. \ref{Fig1}), and mounted on a cold finger thermally anchored to the mixing chamber stage of a dilution refrigerator in cryogenic vacuum. The motion of the fundamental out-of-plane flexural mode of the beam at a frequency $\omega=2\pi\times 17.5$ MHz is actuated and detected with the magnetomotive scheme \cite{cleland_external_1999}, the electromotive signal being detected by a lock-in amplifier [see Fig. \ref{Fig2}a)]. %Lossy coaxial cables down to the 3 K stage are used, followed by superconducting NbTi coaxial cables, and realize together with a 2 k$\Omega$ bias resistor effective low-pass filtering of external radiation, in addition to several shields enclosing the mixing chamber plate. In-situ calibration of applied forces and detected motions taking into account injection and detection losses is realized \cite{collin_-situ_2012}. 
The Al layer (whose critical temperature at zero magnetic field is measured to be $T\approx 1.4$ K) is quenched at all temperatures for in-plane magnetic fields larger than 320 mT, and all the data in the normal state are obtained for fields larger than 600 mT. Damping and frequency shift due to remaining magnetomotive losses \cite{cleland_external_1999} or to intermediate superconducting properties \cite{lulla_evidence_2013} are carefully characterized and subtracted to reveal the intrinsic mechanical damping. For each measurement point in the normal state, we have used low excitation currents to minimize Joule heating. The measurements were repeated over several thermal cycles and found reproducible.%The second sample is a 50 $\mu$m long beam embedded in a superconducting LC resonator \cite{teufel_dynamical_2008}. Sideband driving of the microwave cavity is realized both in red and blue-detuned configuration to characterize and subtract the dynamical backation of the microwave photon field on the resonator that affects the damping rate \cite{aspelmeyer_cavity_2014}. In both cases, the damping and frequency shift due to intrinsic phenomena is monitored as a function of temperature. The results are shown in Fig. \ref{Fig2}.

%Let us focus first on the 15 $\mu$m sample.
We observe a logarithmic dependence of the frequency shift $\delta\omega_0/\omega_0=C\ln(T/T_0)$ [with respect to an arbitrary reference $\omega_0$, see Fig. \ref{Fig2}b)] below 1 K, which we interpret as evidence for TLS-driven behavior \cite{phillips_tunneling_1972}, in particular since $C=2.1\times 10^{-5}$ agrees with commonly reported values for similar structures \cite{zolfagharkhani_quantum_2005,hoehne_damping_2010,venkatesan_dissipation_2010}. This behaviour arises from the resonant interaction of applied phonons through magnetomotive driving and TLSs \cite{phillips_tunneling_1972} (while the relaxational interaction, which we address further in this work to interpret the damping measurements, yields a $T^6$ dependence at low temperature \cite{fefferman_elastic_2010}, which is negligible compared to the resonant contribution to the frequency shift). The frequency shift above 1 K follows a $T^3$ law which we attribute to thermal expansion mismatch between the two layers \cite{collin_metallic_2010}. Meanwhile, the measured damping rate represented in Fig. \ref{Fig3} is divided into two regimes in temperature: above 1-2 K it essentially reaches a nearly constant plateau around $1.2$ kHz. Below 600 mK at low magnetic field (metallic layer in the superconducting state), it decreases linearly with temperature. Below typically 70 mK, the damping rate and frequency shift measurements were rendered complicated by both frequency noise (which is no longer negligible compared to the typical damping rate) and by the low signals measured as a result of a low ($<100$ mT) magnetic field application. This explains why the dispersion on the data and the error bars are significant at 30 mK.

Switching the metallic layer to the normal state with the magnetic field leads to a strikingly different dissipation rate [see Fig. \ref{Fig3}, blue dots] below $700-800$ mK, while it does not contribute further to the resonance frequency shift within our experimental accuracy (see Fig. \ref{Fig2}b) down to roughly 200 mK. Below this temperature, the frequency shift in the normal state slightly deviates from the logarithmic trend. Those observations complement previous results \cite{lulla_evidence_2013} obtained with a low-stress goalpost-shaped silicon nanoresonator (see Fig. 3 inset), with very similar features. This suggests that the mechanism at stake in the normal state is independent of geometry or mechanical properties. Between roughly 150\,mK and 1 K the damping in the normal state
shows a sublinear power law-like behavior and reaches a saturation threshold for lower temperatures, consistently with previous measurements in similar conditions \cite{Imboden_2009,zolfagharkhani_quantum_2005,venkatesan_dissipation_2010}
where $T^\alpha,\alpha=0.3-0.7$ dependences were reported. Note that the saturation temperature range roughly coincides with the range where the frequency shift deviates from the logarithmic trend. We ascertain that the extra damping in the normal state, down to 50 mK, is not caused by a mechanical non-linearity: the damping rate remains unaffected by the displacement amplitude within our experimental accuracy \cite{Suppmat}. Additionally, it does not depend on the current levels injected \cite{Suppmat} in our experimental range (up to $\sim$ 70 nA), consistently with \cite{lulla_evidence_2013}. This rules out Joule dissipation due to the driving scheme as a mechanism of damping saturation as well as a significant cause of thermal decoupling. 
\section{Interpretation in the superconducting state}
The temperature dependence in the superconducting state can be explained using the canonical STM appended with NEMS constraints such as low-dimensionality \cite{Hauer2018,Behunin2016}. In addition, the high tensile stress is shown to quantitatively affect the magnitude of TLS-induced damping \cite{MailletThesis2018,Suppmat} (but not the NEMS elastic constants). Let us consider a TLS inside the structure with asymmetry $\Delta$ and tunneling amplitude $\Delta_0$ between the two potential wells (see Fig. \ref{Fig1}). Considering only the local ground state of each well (the first excited state, of energy comparable with Debye energy, is irrelevant at low temperatures), the bare Hamiltonian of a single TLS is analogous to that of a 1/2 spin. It writes in the TLS position basis corresponding to left and right well locations: $\mathcal{H}_0=1/2(\Delta_0\hat{\sigma}_x+\Delta\hat{\sigma}_z)$, where $\hat{\sigma}_{x,z}$ are Pauli matrices. The TLS eigenstates energies are then easily obtained $\varepsilon_{\pm}=\pm 1/2\sqrt{\Delta^2+\Delta_0^2}$, with corresponding wavefunctions that are delocalized over the two wells. Following the usual STM framework, the NEMS is put into an oscillating motion at angular frequency $\omega$, and the structure undergoes an axial oscillating strain $\mathcal{E}$. The strain field changes the local potential energy landscape that defines the TLS, leading to a modulation of its energy splitting $\varepsilon = \sqrt{\Delta^2+\Delta_0^2}$, formalized by the coupling Hamiltonian in the TLS position basis $\hat{\mathcal{H}}_{int}=\gamma\mathcal{E}\hat{\sigma}_z$, with $\gamma=\frac{1}{2}\partial\Delta/\partial\mathcal{E}$ the TLS-strain coupling strength and $\hat{\sigma}_z$ the TLS diagonal Pauli matrix. Subsequently, the TLS returns to equilibrium by exchanging energy with the phonon bath (the thermal strain field) following the same coupling Hamiltonian over a characteristic time $\tau$. This causes a lagging stress response, leading to mechanical energy dissipation. The power dissipated per unit volume $V$ writes: 
\begin{multline}
\label{Pvol}
P_V=\frac{P_0\omega(\gamma\mathcal{E}_0)^2}{2}\times\\\int_{0}^{\infty}\mathrm{d}u\int_0^1\mathrm{d}v\frac{v^2}{1-v^2}\,\mathrm{sech}^2\left(\frac{u}{2}\right)\frac{\omega\tau(u,v,T)}{1+\omega^2\tau^2(u,v,T)},
\end{multline}
with $\mathcal{E}_0$ the strain oscillation amplitude and $P_0$ the TLS density of states per unit volume. For a string undergoing flexure \cite{cleland_foundations_2002}, the macroscopic imposed strain field oscillates with an amplitude
$\mathcal{E}_0\propto\frac{\partial^2\Psi(z)}{\partial z^2}x_0$, where $x_0$ is the NEMS oscillation amplitude at mid abscissa and $\Psi(z)$ is the excited mode shape. For a high-stress doubly clamped beam, the fundamental flexural mode shape writes: $\Psi(z)=\cos(\pi z/l)$ with $z=0$ at mid abscissa of the beam. The total power $P$ is first obtained through integration of $P_V$ in Eq. (\ref{Pvol}) over the NEMS dimensions \cite{Suppmat}. It is then independently obtained by macroscopic arguments, using the mode dynamics equation: $P=\frac{1}{2}m\omega^2x_0^2\Gamma$, where $m=\rho ew\int\Psi^2(z)\mathrm{d}z=\rho ewl/2$ is the effective mass of the NEMS in its fundamental out-of-plane flexural mode, $\rho$ being its mass density. By equating the two expressions, we thus obtain the generic expression of the TLS ensemble contribution to the NEMS damping rate resulting from the strain modulation \cite{Suppmat}:
\begin{multline}
\label{damping_rate_th}
       \Gamma[\tau] = C\omega\iint\mathrm{d}\varepsilon\mathrm{d}\Delta\times
       \\
       P(\varepsilon,\Delta)\left(\frac{\Delta}{\varepsilon}\right)^2\mathrm{sech}^2\left(\frac{\varepsilon}{2k_BT}\right)\frac{\omega\tau(\varepsilon,\Delta)}{1+\omega^2\tau^2(\varepsilon,\Delta)},
\end{multline}
Here we introduce the commonly assumed \cite{phillips_tunneling_1972} TLS distribution $P(\varepsilon,\Delta)=\varepsilon/(\varepsilon^2-\Delta^2)$ and $C=P_0\gamma^2\pi^2e^2/12\sigma_0l^2$, with $\sigma_0=1.1\pm 0.1$ GPa  the in-built axial stress. Note that $C$ is theoretically identical to the logarithmic slope of the frequency shift \cite{Phillips_1987}, but its expression differs from the one commonly derived \cite{Phillips_1987} due to the influence of high tensile stress on the excited flexural mode \cite{Suppmat}. For high enough temperatures, the damping reaches a plateau $\pi\omega C/2$, regardless of microscopic TLS scattering mechanisms. From the measured plateau we extract $C=4.3\times 10^{-5}$, in qualitative agreement but differing by a factor of 2 from the value given by the frequency shift measurement. Similar discrepancies have been reported \cite{zolfagharkhani_quantum_2005,hoehne_damping_2010,venkatesan_dissipation_2010,lulla_evidence_2013} and may be due to our simplified model that neglects e.g. the bilayer structure or inhomogeneities in the TLS distribution.
 %Writing the TLS susceptibility following Philips \cite{phillips_tunneling_1972} $\chi_t\propto\left(1+i\omega\tau\right)^{-1}$, two regimes are noteworthy: The strain-TLS coupling Hamiltonian writes in the TLS energy eigenbasis $\mathcal{H}_{int}=\frac{\gamma\mathcal{E}}{\varepsilon}\left(\Delta\sigma_z+\Delta_0\sigma_x\right)$. We have derived a renormalized glassy range plateau value which writes as a function of the bulk "glass constant" $C_{bulk}=P_0\gamma^2/(\rho c^2)$:
	%\begin{equation}
	%C_r=\left(\frac{\pi e}{l}\right)^2\cdot\frac{E}	%	{12\sigma_0}C_{bulk}
	%\end{equation}
		%We notice that the beam value is essentially renormalized both by the aspect ratio and the stored in-built stress. In the bending limit $EI\gg\sigma_0A$, we recover $C_r=C_{bulk}$ because the dispersion relation $\omega\propto k^2$ appearing through the strain field expression compensates the renormalization.
\begin{figure}
	\centering
	\includegraphics[width=8.6cm]
	{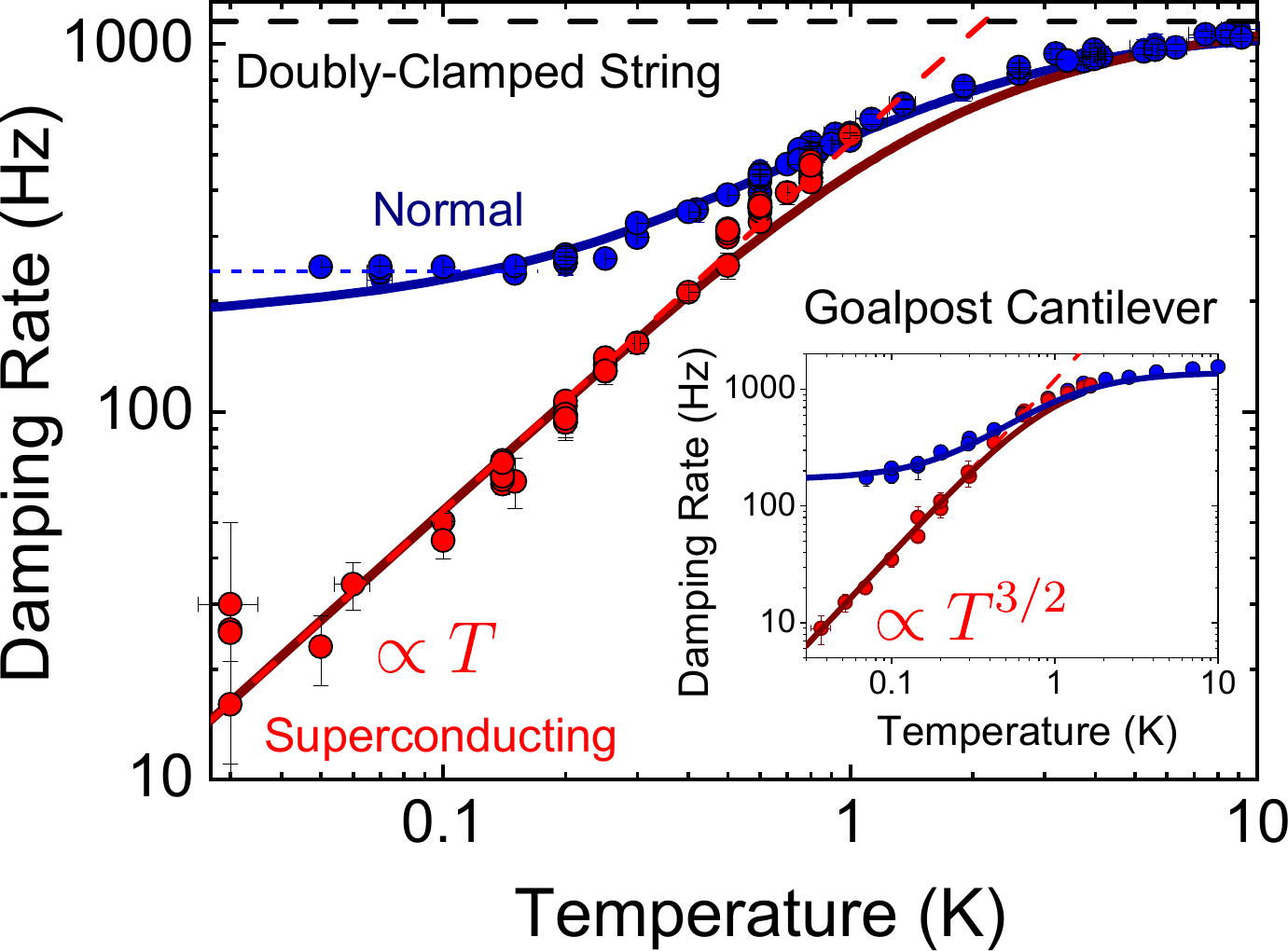}
	\caption{NEMS damping rate as a function of refrigerator's base temperature in normal (blue) and superconducting (red) states, with the magnetomotive contribution subtracted. Inset: data from Ref. \cite{lulla_evidence_2013}. Solid curves (main and inset) are fits using Eq. (\ref{damping_rate_th}) with relaxation rates due to phonons [Eq. (\ref{TLS_relax})] and electrons [Eq. (\ref{relax_electrons})] to electrons. Dashed red and black lines correspond to low and high temperature asymptotic behaviors, respectively. The blue dotted line indicates a possible saturation above 100 mK due to thermal decoupling (see text).}
	\label{Fig3}
\end{figure}

Below 600 mK, the measured linear dependence is consistent with previous reports for similar beams \cite{sulkko_strong_2010, hoehne_damping_2010}, but is in contradiction with the usual $T^3$ dependence. However, at subkelvin temperatures, the dominant phonon wavelength $\lambda = hc/2.82k_BT$ ($\sim 100$ nm at 1 K) becomes bigger than the transverse dimensions of the resonator. As a result TLS relax to equilibrium by exchanging energy only with longitudinal phonon modes \cite{Suppmat}, which are not confined and thus lie at lower energies. They realize a quasi one-dimensional phonon bath with constant density of states $l/\pi c_l$ and linear dispersion relation, as proposed e.g. in Refs. \cite{sulkko_strong_2010,hoehne_damping_2010} and more extensively studied in Refs. \cite{Behunin2016,Hauer2018,MailletThesis2018}. At first non-vanishing order in the thermal strain field perturbation, Fermi's Golden rule yields the relaxation rate of a single TLS due to its interaction with the phonon bath \cite{Suppmat,Hauer2018}:
\begin{equation}
\label{TLS_relax}
\tau^{-1}_{ph}=\frac{\gamma^2}{\hbar^2\rho ewc_l^3}\frac{\Delta_0^2}{\varepsilon}\coth\left(\frac{\varepsilon}{2k_BT}\right),
\end{equation}
%The real and imaginary parts of the TLS susceptibility as a function of the reduced energy splitting $u=\varepsilon/k_BT$ and asymmetry $v=\Delta/\varepsilon$ are plotted in Figure \ref{Fig3}.
%As expected, TLS with vanishing tunneling amplitude $\Delta_0$ dominate the high-temperature (above 1 K) response, because high temperatures destroy the TLS coherence. Accordingly, we see that excitations with larger tunneling amplitude $\Delta_0\sim 0.7\varepsilon$ start to dominate below 1 K.

where $c_l\sim 6000$ m/s and $\rho=2.9\times 10^3$ kg/m$^{3}$ are the longitudinal sound speed and mass density for SiN, respectively. Combining this relaxation rate with the damping expression (\ref{damping_rate_th}), we capture both the damping data in the low temperature range when the field is low enough to maintain superconductivity in the metallic layer, and the high temperature limit. To consistently fit both the low ($\Gamma[\tau_{ph}]\propto \gamma^2CT$, dashed red line in Fig. \ref{Fig3}) and high ($\Gamma[\tau_{ph}]\propto C$) temperature regimes we use $\gamma=9.8$ eV and $P_0=2.2\times 10^{44}$ J$^{-1}$.m$^{-3}$. The fitted interaction energy is rather high (one would expect it more around 1 eV), but may likely reflect a non-uniform distribution of the TLS inside the beam \cite{hamoumi_microscopic_2018}, which is out of the scope of this study. Note that our expression does not fit the data in the superconducting state between 600\,\,mK and 1 K, which we attribute to the substantial density of quasiparticle excitations in this range, that should contribute to an excess damping through the same mechanism as electrons in the normal state \cite{Suchoi_2017}: in fact, above 800 mK the data in both states are identical within experimental accuracy, as observed previously \cite{lulla_evidence_2013}. This shall be addressed elsewhere.

\begin{figure*}[ht]
	\centering
	\includegraphics[width=17.2cm]
	{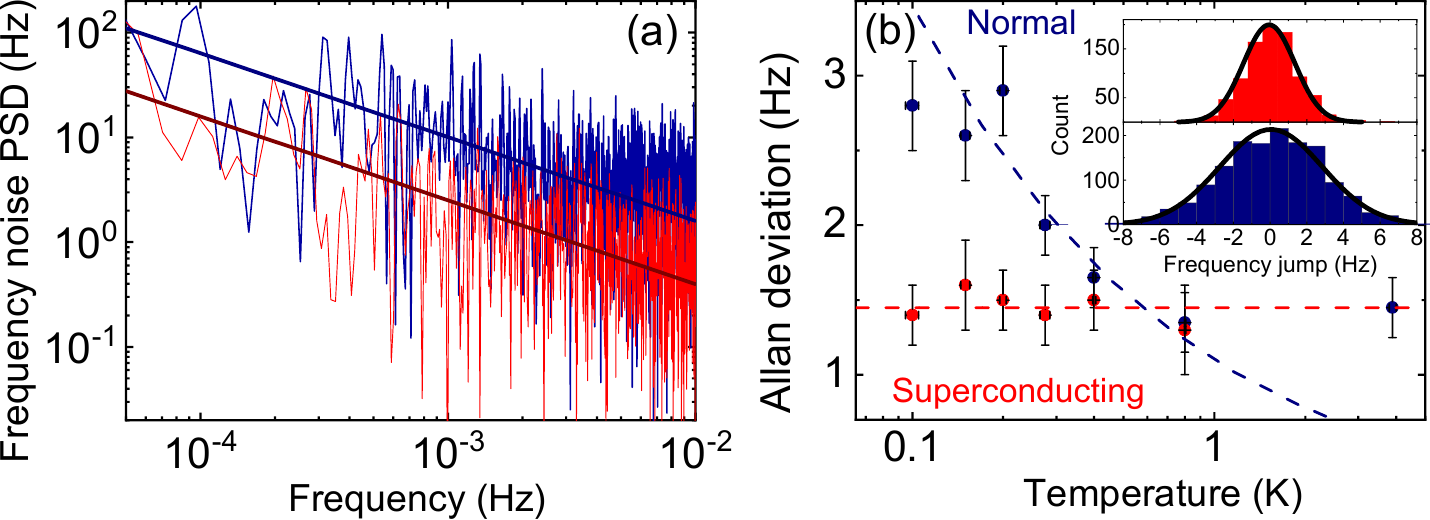}
	\caption{a) Frequency noise spectrum at $T=100$ mK in the superconducting (red) and normal (blue) state of the metallic layer. Solid lines are $1/f^{\mu}$ type functions, with $\mu=0.8\pm 0.2$ here. b) Allan deviation $\sqrt{\langle\delta f^2\rangle}$ of frequency noise ($\delta f$ is the jump between two successive frequency measurements), extracted from the standard deviation of Gaussian fits to the jumps histograms. Dashed lines (constant for the superconducting state, $\propto 1/\sqrt{T}$ for the normal state) are guides to the eye. Inset: comparison of frequency jumps histograms in the superconducting and normal state at $T=100$ mK. Solid lines are Gaussian fits with mean $\langle\delta f\rangle=0$ and standard deviation thus directly reflecting the Allan deviation measured over $\sim 10^4$ s.}
	\label{Fig4}
\end{figure*}

%However, in stark contrast with the 3D standard theory where "thermal TLS" with energy splitting $\varepsilon\sim k_BT$ dominate the relaxation \cite{fefferman_low_2009}, here low-energy TLS $\varepsilon\ll k_BT$ are predominant.
\section{Interpretation in the normal state}
In the normal state, it is natural to consider the conduction electrons as an additional relaxation channel for TLS in parallel with the phonon bath: when a TLS entity tunnels, the Coulomb potential that scatters conduction electrons is modified, which translates as an effective electron-TLS coupling. The effective Hamiltonian writes $\mathcal{H}_{el}=\sum V_{k,k'}\hat{c}_{k'}^{\dagger}\hat{c}_{k}\hat{\sigma}_z$, where $\hat{\sigma}_z$ is the TLS diagonal Pauli matrix, $\hat{c}_k^{(\dagger)}$ the electron creation (annihilation) operator at wavevector $k$ and $V_{k,k'}$ the coupling matrix elements. We further make the assumption that the interaction is uniform near the surface of the Fermi sea, which enables the simplification $V_{k,k'}\approx V$. This is reasonable insofar as electronic excitations are scattered within a bandwidth $k_BT$ very small compared to the Fermi energy. The corresponding relaxation rate is again obtained by Fermi's Golden Rule \cite{Suppmat}:
\begin{equation}
\label{relax_electrons}
\tau_{el}^{-1}=\frac{4\pi K}{\hbar}\frac{\Delta_0^2}{\varepsilon}\coth\left(\frac{\varepsilon}{2k_BT}\right),
\end{equation}
where $K=(n_0V\Omega)^2$ is a normalized electron-TLS coupling strength, $n_0=1.07\times 10^{47}$ J$^{-1}$m$^{-3}$ being the electronic density of states at the Fermi level for aluminum, $V$ an averaged coupling constant, and $\Omega$ the effective interaction volume, which should vanish beyond the metallic screening length ($\lesssim 1$ nm). The energy dependence is the same as in the case of phonon-assisted relaxation, which led previous studies to invoke additional effects \cite{zolfagharkhani_quantum_2005}. However, the averaging over the full TLS distribution leads to a weakened dependence in temperature of the damping rate in the normal state. Indeed, it is certain that not all TLS interact with electrons, due not only to the fact that conduction electrons are located solely in the metallic layer, but also because of the very nature of the TLS-electron interaction: depending on the microscopic nature of the TLS, the interaction may be very weak, as in the case of a mere dislocation in the Al layer \cite{hoehne_damping_2010}. Therefore, electrically "neutral" TLS interact with phonons, but not with electrons, which leads to separate averages over charged and neutral TLS. 

For "neutral" TLS and TLS within the SiN layer, which relax only due to interactions with phonons, the crossover temperature $T^*_{ph}$ between the plateau and the power law regime is defined by the condition $\omega\tau_{ph}\sim 1$, that they must all satisfy. Meanwhile for charged TLS, the crossover temperature $T_{el}^*$ is shifted down by electron-assisted relaxation because for these TLS the condition $\omega\tau\sim 1$, where now $\tau^{-1}=\tau_{ph}^{-1}+\tau_{el}^{-1}$, is modified. Thus, for $T_{el}^*<T<T_{ph}^*$, an intermediate regime emerges. At ultra-low temperatures, the $\propto{T}$ relaxation regime should be recovered when all TLS in the structure satisfy $\omega\tau\gg 1$. We fit the data in the normal state using a balanced expression of the damping rate $\Gamma_N=(1-x)\Gamma[\tau_{ph}]+x\Gamma[(\tau_{ph}^{-1}+\tau_{el}^{-1})^{-1}]$ using for the two contributions the generic form of Eq. (\ref{damping_rate_th}), by assuming a fraction $x=0.17$ of TLS interacting with electrons with a coupling constant $K=0.07$ well below 1, which allows us to neglect Kondo-type strong coupling corrections \cite{coppersmith_low-temperature_1993,leggett_dynamics_1987}, yielding an electron-TLS coupling energy $V$ in the 0.1 eV range. Based on this set of parameters, we evaluate crossover temperatures $T_{ph}^*\approx 1$ K (as visible in Fig. \ref{Fig3}) and $T_{el}^*\approx$ 0.9 mK \cite{Suppmat}, the latter being unreachable with standard dilution refrigeration. For comparison, we have also fit the data obtained on another sample, namely the goalpost-shaped silicon nanocantilever (with dimensions comparable to the SiN beam, 100 nm thick $\times$ 250 nm wide, two $3~\mu$m long feet linked by a $7~\mu$m long paddle), covered with a 50 nm aluminum layer similar to that of the high-stress SiN sample, measured in Ref. \cite{lulla_evidence_2013}. The dissipation in the superconducting state features a $T^{3/2}$ power law at low temperatures which might owe to a non-linear dispersion relation of flexural mechanical modes in the device \cite{seoanez_surface_2008}. The data are reproduced using a semi-phenomenological expression for the TLS-phonon relaxation rate \cite{Suppmat}. The normal state data were fit with parameters $x=0.12$ and $K=0.11$, comparable with the ones used for the high stress SiN sample. This is consistent with a mechanism independent from the mechanical properties of the resonator, as the interaction lengthscale is much smaller than any mechanical dimension.

The model captures the data in the 0.15-1 K range, as seen in Fig. \ref{Fig3}. In particular, it reproduces the sublinear power law-like behaviour consistently reported in previous works \cite{lulla_evidence_2013,zolfagharkhani_quantum_2005,venkatesan_dissipation_2010} in similar experimental conditions. Although our model captures to a large extent the saturation observed below 150 mK and reported in nearly every nanomechanical damping study for the lowest operation temperatures, it is likely that a hot electron effect also causes overheating of the structure \cite{roukes_hot_1985}. This saturation can be estimated through the frequency shift measured in the normal state: it provides a thermometer by using the deviation from the logarithmic shift at the lowest temperatures. This could mean that the electron temperature does not go lower than 100 mK at the lowest refrigerator temperature (50 mK) in the normal state [see Fig. \ref{Fig2}b), blue dots]. This is consistent, within experimental accuracy, with the observed damping saturation slightly above the theoretical fit below 100 mK [see Fig. \ref{Fig3}, blue dotted line]. We could attribute these two features to thermal decoupling caused by parasitic radiation in the 10 pW range, which heats up the Aluminum layer, i.e., the electron bath to which the TLS ensemble (and ultimately, then, the mechanical mode) thermalize \cite{Suppmat}. Ultimately, we see evidence, supported by our analysis, that both thermal decoupling and the TLS-induced damping contribute to the damping saturation in the normal state at the lowest operation temperatures, but more work is needed to separate the two contributions.
%Note that this saturation was reported only for low-dimensional structures: this can be explained by the fact that around 100 mK, in most metals the phonon bath becomes one dimensional for such structures, which leads to poorer electron-phonon heat flow, enhancing the hot electron effect.   

The proposed modeling is fairly generic, and makes minimal assumptions on the microscopic nature and location of TLS. An educated guess based on our results would locate those TLS which interact with electrons at the interfaces, between the SiN and Al layer, and between the Al layer and its native oxide at the NEMS surface: indeed, tunneling atoms are less likely to exist within the metallic layer due to long-range order, leaving kinks on dislocations, which are only weakly interacting with electrons, as most probable candidates for TLS in polycrystalline aluminum, as proposed in Ref. \cite{hoehne_damping_2010,fefferman_elastic_2010}. 

Note that a recent study with similar methods \cite{kamppinen2021dimensional} but bare aluminum resonators reports very little difference between damping in the normal and superconducting state, with a much smaller interaction constant $V\sim 10^{-4}$ eV. This may be seen as further evidence that aluminum does not itself possess defects that would act as strongly interacting TLS, and also points towards a location of interacting TLS at the Al-SiN interface rather than in the Al native oxide. %Finally, note that the observed saturation at the lowest operation temperatures (as well as the absence of excess damping in Ref. \cite{kamppinen2021dimensional} with bigger metallic layer dimensions) also rule out a mechanism based on electron-phonon coupling, that would feature a strong temperature dependence \cite{LindenfeldPRB}.
% Note that these kinks may have a distribution in energies which favors more symmetric TLS configurations \cite{Fefferman2010PRB}, unlike the one we have used in our modeling.

As an opening for further investigations, we have measured the resonant frequency noise of our device using the dynamical bifurcation properties of the NEMS in the Duffing regime \cite{aldridge_noise-enabled_2005,Maillet2018}. The observed spectrum is that of a $1/f$-type noise typical of a collection of switching two-level systems \cite{dutta_low-frequency_1981}. Notably, a visible increase of its magnitude is observed below 1 K when the metallic layer is switched to the normal state, as seen in Fig. \ref{Fig4}. Since at these temperatures the switching, which is due to tunneling, is mainly induced by TLS-electrons interactions, it is reasonable to expect that electron-TLS interactions cause the excess frequency noise: the tunneling events occurring during TLS relaxation cause local rearrangement of atoms, and may thus lead to stress  (i.e., frequency) fluctuations.

In conclusion, our results support the idea that electron-driven TLS relaxation in metallic nanomechanical structures is the dominant mechanism of damping, through timescale decoupling between phonon- and electron-induced TLS relaxation. This may bring an answer to several issues raised over the last two decades by nanomechanical damping measurements at low temperatures. In addition, we expect that measurements of frequency noise may shed further light on microscopic mechanisms at work, possibly highlighting interactions between TLS \cite{Fefferman2008}, through e.g. a careful extraction of the exponent of frequency noise \cite{Faoro2015}.

%%%%%%%%%%%%%%%%%%%%%  Experimental setup
%%%%%%%%%%%%%%%%%%%%%%%%%%%%%%%%%%%%%%%%%

% 1 sentence summary of fabrication, in particular if Applied Physics Letter
%

%%%%%%%%%%%%%%%%%%%%%%%%%%%%%%%%%%  Fig2
%%%%%%%%%%%%%%%%%%%%%%%%%%%%%%%%%%%%%%%%%
% transmission and backward transmission of 2 single-MO amplifiers
% + their coherent sum

%%%%%%%%%%%%%%%%%%%%%%%%%%%%%%%%%%  Fig3
%%%%%%%%%%%%%%%%%%%%%%%%%%%%%%%%%%%%%%%%%
% 4 S-parameters

\begin{acknowledgments} We acknowledge the use of the N\'eel facility {\it Nanofab} for the device fabrication.
We acknowledge support from the 
ERC CoG grant ULT-NEMS No. 647917, 
StG grant UNIGLASS No. 714692.
The research leading to these results has received funding from the European Union's Horizon 2020 Research and Innovation Programme, under grant agreement No. 824109, the European Microkelvin Platform (EMP).
\end{acknowledgments}

\bibliography{damping_paper_arxiv}
\bibliographystyle{ieeetr}

\onecolumngrid
\appendix
\setcounter{figure}{0}
\renewcommand{\thepage}{S\arabic{page}} 
\renewcommand{\thesection}{S\arabic{section}}  
\renewcommand{\thetable}{S\arabic{table}}  
\renewcommand{\thefigure}{S\arabic{figure}} 

 \section{Overheating in the normal state}
Since the driven mechanical mode dissipates energy mainly through the TLS ensemble, which is itself coupled to the electron bath of temperature $T$, it is natural to think that any source of overheating of the electrons can lead to thermal decoupling of the mechanical mode from the cryostat temperature $T_0$. Based on the logarithmic trend observed in Fig. 2 of the main text, which, within our resolution, is identical for both layer states down to 200 mK, we assume that electron thermalization is ensured down to 200 mK.

Let us suppose that a heat load per unit volume $\dot{q}_{ext}$ affects the metallic layer of the beam. We assume that the beam is thermalized at both its ends to the pads, ie $T(z=\pm l/2)=T_0$. In addition, the symmetry of the problem imposes that $\partial T/\partial z|_{z=0}=0$, ie that the beam temperature is maximal at its mid abscissa. The temperature profile of the beam is then determined by the heat diffusion equation:
\begin{equation}
\label{diff_beam}
C_V\frac{\partial T}{\partial t}=\frac{\partial}{\partial z}\left(\kappa(T)\frac{\partial T}{\partial z}\right)+\dot{q}_{ext}-\dot{q}_{eph},
\end{equation}

where $C_V$ is the volumic specific heat, $\kappa$ is the Al thermal conductivity, and where for completeness we consider the volumic heat flow of electrons to the phonon bath (assumed at temperature $T_0$), $\dot{q}_{eph}=\Sigma(T^5-T_{0}^5)$, with $\Sigma\approx 3\times 10^8~\mathrm{W}\cdot\mathrm{K}^{-5}\cdot\mathrm{m}^{-3}$ the aluminum electron-phonon coupling constant \cite{meschke_electron_2004}. Using Wiedemann-Franz law, we write the thermal conductivity $\kappa=\pi^2k_B^2T/3\rho_e e^2$, $\rho_e$ being the resistivity of the aluminum coating of the NEMS.

We consider two sources of external heating: first, electrons are heated up due to an external power $\dot{Q}_0$, presumably radiated from upper stages. Second, Joule dissipation $\dot{Q}_J=RI_{ac}^2[1+\cos(2\omega t)]/2$ occurs in the beam coating, due to the current necessary to the magnetomotive drive and detection. Here $I_{ac}$ is the current drive amplitude, and $R\approx 100~\Omega$ is the resistance of the NEMS aluminum layer. Considering only the time-averaged profile, we rewrite the heat diffusion equation:

\begin{equation}
\label{diff_beam_DC}
-\frac{l^2\mathcal{L}}{R}\frac{\partial}{\partial z}\left(T\frac{\partial T}{\partial z}\right)=\frac{RI_{ac}^2}{2}+\dot{Q}_{0}-\Sigma V_{Al}(T^5-T_{0}^5),
\end{equation}

with $\mathcal{L}=\pi^2k_B^2/3e^2\approx 2.44\times 10^{-8}~\mathrm{V}^2/\mathrm{K}^2$ the Lorenz number and $V_{Al}=30$ nm $\times 300$ nm $\times 15~\mu$m the NEMS metallic layer nominal volume. Numerically solving Eq. (\ref{diff_beam_DC}), we can make the following observation: to obtain the temperature $\bar{T}$ averaged over the beam's profile that accounts for the deviation of the frequency shift from the logarithmic trend, a heat leak $\dot{Q}_0\approx 11$ pW must be introduced (see Fig. \ref{T_decouple}). Meanwhile, the corresponding electron-phonon heat flow is at most 20 fW below 200 mK, which allows us to safely neglect it. In addition, the maximum currents used to obtain the data presented in the main text are of the order of 30 nA, leading to about 100 fW of Joule power dissipated, which is again too small to explain the observed decoupling. 
\begin{figure}[ht]
	\centering
	\includegraphics[width=12cm]
	{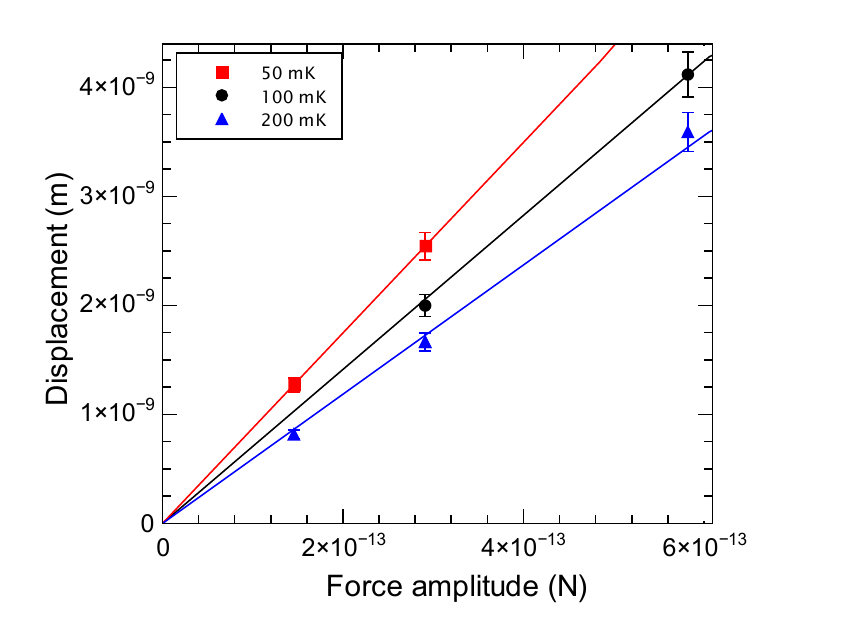}
	\caption{Maximum amplitude displacement of the NEMS beam vs applied force (peak value), at magnetic field $B=0.87 $T, for three different temperatures (50, 100, 200 mK). The currents corresponding to the three forces used are 17.5, 35 and 70 nA.}
	\label{xvsF_SI}
\end{figure}

We can further establish experimentally that the drive and detection scheme plays no significant role in thermal decoupling, by driving the NEMS at a given cryostat temperature with different force (and therefore, current) amplitudes, which translates as different Joule powers dissipated. Even if the NEMS enters the non-linear Duffing regime, we have access to the NEMS damping rate by using the relation between displacement amplitude and applied force at the frequency corresponding to the maximum vibration amplitude $x_{max}=F/m\omega\Gamma$. As shown in Fig. \ref{xvsF_SI}, this linear relation (with no offset) holds at all powers injected at cryostat temperatures 50, 100 mK and 200 mK. This implies that the damping rate is unaffected by Joule dissipation within our experimental accuracy, and in turn that this dissipation does not cause significant thermal decoupling.

To conclude, based on our analysis, we could attribute the thermal decoupling of the mechanical mode predominantly to a parasitic heat leak of about 11 pW. This heat leak may originate from the thermal radiation of the coil encircling the sample (which had no further shield) and anchored to a screen thermalized to the still plate (at about 800 mK), or from imperfections in the drive and detection lines filtering, causing electrical noise from hotter parts of the refrigerator to flow in the structure. Regardless of its origin, such a heat leak causes significant ($> 10$ \%) thermal decoupling below 200 mK that is consistent with the deviation of the frequency shift data in the normal state from the logarithmic trend observed in the superconducting state at all temperatures (and in the normal state above 200 mK). It also is consistent with the saturation of the NEMS damping rate observed below 150-200 mK, although the temperature dependence of this damping rate predicted by our model (see below) is anyway expected to be modest in this temperature range. Therefore, the two possible causes of damping saturation in this temperature range (TLS relaxation and overheating) can hardly be distinguished from one another. Note, however, that the impact of overheating can safely be neglected above 200 mK.
\begin{figure}[ht]
	\centering
	\includegraphics[width=10cm]
	{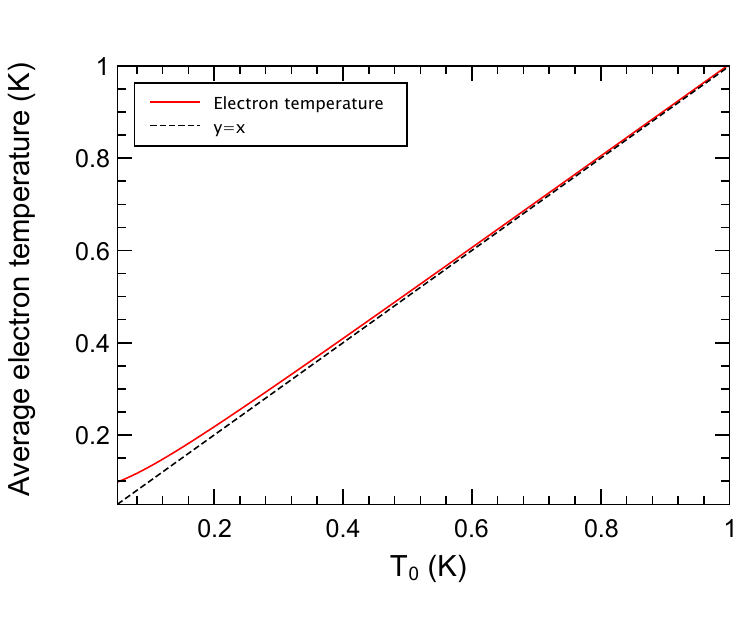}
	\caption{Temperature of the electrons averaged over the NEMS structure, assuming a parasitic heat input $\dot{Q}_0=11$ pW. At a cryostat temperature $T_0=50$ mK, the electrons are overheated to $\approx 100$ mK.}
	\label{T_decouple}
\end{figure}
 \section{Relaxational TLS-strain interaction}
The following derivation is adapted from several reference texts \cite{Phillips_1987}, which we briefly recall while adding features such as the reduced dimensionality (phonon-wise) and the string nature of the nanomechanical beam. A single TLS (see Fig. 1 of the main text) is modelled as a double-well potential, each one locally in its ground state (the intra-well energy spacing is on the order of the Debye frequency $\gg k_BT/\hbar$). The double well is characterized by its asymmetry $\Delta$, barrier height $V_0$ and distance $d$ in space, which is typically a few \r{A}. Following the WKB approximation we model the overlap between the two wells' wave functions $|\Psi_L\rangle,|\Psi_R\rangle$ to vanish exponentially as $\langle\Psi_L|\Psi_R\rangle= e^{-\lambda_G}$ where $\lambda_G=d\sqrt{2m_aV_0}/\hbar$ is the Gamov factor. The coupling matrix element between the two wells write $\Delta_0=\hbar\Omega_0e^{-\lambda_G}$, where $\Omega_0$ is a reference energy scale. The Hamiltonian of a single TLS in the position basis $\{|\Psi_L\rangle,|\Psi_R\rangle\}$ thus writes:
\begin{equation}
\label{TLS_ham}
\hat{\mathcal{H}}_0^{(p)}=\frac{1}{2}\begin{pmatrix}
\Delta & \Delta_0 \\
\Delta_0 & -\Delta \\ 
\end{pmatrix}=\frac{\Delta_0}{2}\hat{\sigma}_x+\frac{\Delta}{2}\hat{\sigma}_z,
\end{equation}
where we introduce the usual Pauli matrices $\sigma_{x,z}$. Eigenenergies $\varepsilon_{\pm}=\pm\sqrt{\Delta^2+\Delta_0^2}/2$ are straightforwardly obtained, and we can thereafter characterize the TLS through its energy splitting $\varepsilon=\varepsilon_+-\varepsilon_-=\sqrt{\Delta^2+\Delta_0^2}$. In typical amorphous solids at low temperatures compared to the glass transition temperature, it is commonly assumed \cite{Phillips_1987} that the distribution of TLS follows $P(\Delta,\Delta_0)=P_0/\Delta_0$. This rewrites for a different set of variables using the jacobian transform $P(\varepsilon,\Delta_0)=P_0/(\Delta_0\sqrt{1-\Delta_0^2/\varepsilon^2})$. Let us now assume the NEMS is driven: this corresponds to a strain perturbation which, at the level of the TLS, influences its energy landscape through local atomic displacements. The applied strain field $\mathcal{E}$ in the MHz range corresponds to a wavelength much larger than the typical TLS interwell spacing $d$ and therefore can be treated as a linear perturbation to the TLS Hamiltonian, which changes the asymmetry of the TLS but negligibly affects its tunneling amplitude. In the position basis, we thus write it $\hat{\mathcal{H}}_{int,ph}^{(p)}=+\gamma\mathcal{E}\hat{\sigma}_z$, where $\gamma=\frac{1}{2}\partial\Delta/\partial\mathcal{E}$ is the deformation potential energy, i.e. the TLS-strain coupling constant. Rewritten in the TLS energy basis $\{|-\varepsilon/2\rangle,|\varepsilon/2\rangle\}\equiv\{|g\rangle,|e\rangle\}$ the interaction Hamiltonian is:
\begin{equation}
\label{TLS_ph_pert}
\hat{\mathcal{H}}_{int,ph}^{(\varepsilon)}=-\frac{\gamma\mathcal{E}}{\varepsilon}\left(\Delta\hat{\sigma}_z-\Delta_0\hat{\sigma}_x\right).
\end{equation}
We then introduce the occupation probabilities of the TLS $p_{g,e}$ and its polarization $p=p_e-p_g$. At thermal equilibrium $p_0=\langle\hat{\sigma}_z\rangle_T=-\tanh(\varepsilon/2k_BT)$. Applying a classical strain field $\mathcal{E}=\mathcal{E}_0\cos(\omega t)$ modulates the energy splitting and thus leads to a change in polarization $\delta p=p-p_0$ away from equilibrium. For small perturbations, one can write the evolution of $p$ using the relaxation time approximation:
\begin{equation}
\label{RT_approx}
\dot{p}=-\frac{p-p_{st}}{\tau},
\end{equation}
where $p_{st}$ is the instantaneous equilibrium value corresponding to a given strain magnitude $\mathcal{E}$, and $\tau$ the relaxation time of the TLS. Developing at first order in the strain field, one can link $p_{st}$ to the thermal equilibrium polarization $p_0$:
\begin{equation}
\label{p_stat}
p_{st}\approx p_0+\frac{\partial p_0}{\partial\Delta}\cdot\frac{\partial \Delta}{\partial\mathcal{E}}\mathcal{E}=p_0+\frac{\gamma\mathcal{E}}{k_BT}\frac{\Delta}{\varepsilon}\mathrm{sech}^2\left(\frac{\varepsilon}{2k_BT}\right).
\end{equation}
Eq. (\ref{RT_approx}) can now be rewritten for $\delta p$:
\begin{equation}
\label{RT_approx_deltap}
\tau\dot{\delta p}=-\delta p-\frac{\gamma\mathcal{E}}{k_BT}\frac{\Delta}{\varepsilon}\mathrm{sech}^2\left(\frac{\varepsilon}{2k_BT}\right).
\end{equation}
Assuming a sinusoidal excitation at frequency $\omega$, i.e. $\mathcal{E}=\mathcal{E}_0e^{i\omega t}$, one can define a TLS susceptibility in polarization $\chi(\omega)=\delta p/\mathcal{E}_0$:
\begin{equation}
\label{TLS_suscept}
\chi(\omega)=\frac{\gamma\mathcal{E}}{k_BT}\frac{\Delta}{\varepsilon}\mathrm{sech}^2\left(\frac{\varepsilon}{2k_BT}\right)\frac{1}{1-i\omega\tau}.
\end{equation}
From this we can write the power dissipated by the applied strain through a single TLS relaxation: using the interaction Hamiltonian (\ref{TLS_ph_pert}), the applied strain changes the internal energy by an amount $\delta U(t)=\gamma\Delta\delta p(t)/\varepsilon$. The imaginary part in frequency domain of the energy change, proportional to that of the TLS susceptibility (\ref{TLS_suscept}), corresponds to the dissipated energy, from fluctuation-dissipation theorem. Integrating over an oscillation cycle, we thus obtain the power dissipated by a continuously strain-driven TLS:
\begin{equation}
P_{\varepsilon}=\frac{(\gamma\mathcal{E}_0)^2}{2k_BT}\left(\frac{\Delta}{\varepsilon}\right)^2\mathrm{sech}^2\left(\frac{\varepsilon}{2k_BT}\right)\frac{\omega^2\tau}{1+\omega^2\tau^2}.
\end{equation}
We then introduce the reduced variables $u=\varepsilon/k_BT,v=\Delta/\varepsilon$. The power dissipated by a collection of TLS per unit volume thus writes
\begin{equation}
\label{Pvol}
P_V=\frac{P_0\omega(\gamma\mathcal{E}_0)^2}{2}\underbrace{\int_{0}^{\infty}\mathrm{d}u\int_0^1\mathrm{d}v\frac{v^2}{1-v^2}\,\mathrm{sech}^2\left(\frac{u}{2}\right)\frac{\omega\tau(u,v,T)}{1+\omega^2\tau^2(u,v,T)}}_{\displaystyle\mathcal{I}(T)}.
\end{equation}
Note that the integral $\mathcal{I}(T)$ diverges in the general case. However, the explicit dependence on $u,v$ of the TLS relaxation time usually regularizes the integrand.
To obtain the total power dissipated we sum over the whole NEMS space. However, care must be taken because the strain is not uniform for our structure, a point which is not usually considered in the standard approach for TLS-driven dissipation but was addressed by Ref. \cite{unterreithmeier_damping_2010}. Note that this is different from considering the contribution of string elongation to energy loss, which can be neglected \cite{unterreithmeier_damping_2010}. For a string undergoing flexure \cite{cleland_foundations_2002}, the macroscopic imposed strain field oscillates with an amplitude
\begin{equation}
\label{strain}
\mathcal{E}_0(x_t,z)=\frac{\partial^2\Psi(z)}{\partial z^2}x_0x_t,
\end{equation}
where $x_0$ is the NEMS oscillation amplitude at mid abscissa, $x_t$ the local coordinate spanning the beam thickness from $-e/2$ to $e/2$ and $\Psi(z)$ is the excited mode shape. For a high-stress doubly clamped beam, the fundamental flexural mode shape writes: $\Psi(z)=\cos(\pi z/l)$. The total power is thus $w\times\int\mathrm{d}z\mathrm{d}x_tP_V(x_t,z)$. We equate it with the power obtained by macroscopic arguments from Newton's second law, $\frac{1}{2}m\omega^2x_0^2\Gamma$, where $m=\rho ewl/2$ is the mode mass of the NEMS, $\rho$ being its mass density. We finally obtain the generic expression of the TLS contribution to the NEMS damping rate:
\begin{equation}
\label{TLS_damping_generic}
       \Gamma[\tau] = C\omega\iint\mathrm{d}\varepsilon\mathrm{d}\Delta
       P(\varepsilon,\Delta)\left(\frac{\Delta}{\varepsilon}\right)^2\mathrm{sech}^2\left(\frac{\varepsilon}{2k_BT}\right)\frac{\omega\tau(\varepsilon,\Delta)}{1+\omega^2\tau^2(\varepsilon,\Delta)},
\end{equation}
where $C=P_0\gamma^2\pi^2e^2/12\sigma_0l^2$, $\sigma_0=1.1\pm 1$ GPa being the pre-stress in the SiN layer, the prefactor $P_0$ of the TLS distribution being absorbed in $C$. $C$ can be seen as a dimensionless TLS-strain coupling strength. In particular, at high enough temperature, the damping rate (\ref{TLS_damping_generic}) reaches a plateau, regardless of the microscopic TLS relaxation mechanism:
\begin{equation}
\label{TLS_damping_plateau}
\Gamma\underset{\omega\tau\ll 1}\approx\frac{\pi\omega C}{2}.
\end{equation}
Interestingly, the $C$ value obtained from the plateau in damping has been experimentally found to vary only little ($10^{-4}-10^{-3}$) for a wide range of macroscopic amorphous solids, which has led to a debate on a possible universality of glasses at low temperature \cite{pohl2002,hunklinger_chapter_1986}. Note that since $\omega\propto\sqrt{\sigma_0}$ and $C\propto 1/\sigma_0$, in the high-stress limit the damping will essentially follow $\Gamma\propto 1/\sqrt{\sigma_0}$, that is, an axial stress load reduces the dissipation induced by TLS. 

Let us also mention that our derived value of $C$ differs from the usual expression $C_0=P_0\gamma^2/\rho c^2$, where $c=\sqrt{E/\rho}$ is the typical sound speed in the material, $E$ being the material Young's modulus. However, we stress that the discrepancy is due to the string nature of the high-stress mechanical resonator combined with its high aspect ratio. The link between our value and the commonly derived one $C_0$ is:
\begin{equation}
C=\left(\frac{\pi e}{l}\right)^2\frac{E}{12\sigma_0}C_0.
\end{equation}
This result is quite interesting: it directly links the damping rate to the aspect ratio of the string, quadratically, as compared with a bulk macroscopic body. 

%Another interesting feature is that in the low-stress limit, the usual expression $C_0$ is recovered, regardless of the aspect ratio. Using the fact that $\omega\approx\sqrt{Ee^2/12\rho}(\pi/2l)^2$ in the low stress approximation, and that for pure beam flexure $\partial^2\Psi(z)/\partial z^2=(\pi/l)^2\Psi(z)$, one can again combine Eqs. (\ref{Pvol}) and (\ref{strain}) to obtain Eq. (\ref{TLS_damping_generic}) with $C_0$ replacing $C$.

In the low temperature limit, TLS typically relax to equilibrium on a timescale longer than the driving period, i.e. $\omega\tau\gg 1$. In that limit, the damping rate is an explicit function of $\tau$ and therefore depends on the microscopic TLS relaxation mechanisms:
\begin{equation}
\label{Damping_lowT_functional_1}
\Gamma\underset{\omega\tau\gg 1}\approx C\int_0^{\infty}\mathrm{d}u\int_0^1\mathrm{d}v\frac{v^2}{1-v^2}\mathrm{sech}^2\left(\frac{u}{2}\right)\tau_{ph}^{-1}(u,v,T),
\end{equation}
We first consider relaxation only to a phonon bath. The generic relaxation rate can be obtained using perturbation theory in the interaction Hamiltonian (\ref{TLS_ph_pert}), where now $\mathcal{E}$ is the strain operator relative to all phonon modes of wavevector $\textbf{k}$ and polarization $s$. We make a few assumptions: first, the beam confines transverse modes, such that their lowest energies are $\hbar c/w,\hbar c/e\gg k_BT$ below 1 K (this is equivalent to saying that the dominant thermal wavelength is larger than these modes' wavelengths). As a result, these modes are not thermally excited and thus will not contribute to the TLS relaxation. Second, since shear produces a much smaller deformation locally than a longitudinal strain, we assume that torsional modes, that emerge from shear, couple much less to TLS than longitudinal modes do. This leaves only longitudinal modes and the two flexural families. Let us address first the longitudinal modes. In a second quantized form \cite{ashcroft_solid_1976}:
\begin{equation}
\label{strain_operator}
\left|\hat{\mathcal{E}}\right|=\sum_{k}\sqrt{\frac{\hbar}{2\rho V\omega_{k}}}k\left(\hat{a}_{k}^{\dagger}+\hat{a}_{-k}\right).
\end{equation}
The matrix element $Q_{k}^{em}(\Delta_0,\varepsilon)$ for such an interaction, considering a TLS emission event into the phonon mode, is
\begin{equation}
\label{TLS_ph_matrix_el}
|Q_{k}^{em}(\Delta_0,\varepsilon)|=\left|\left\langle g,n_{k}+1\left|\hat{\mathcal{H}}_{int,ph}^{(\varepsilon)}\right|e,n_{k}\right\rangle\right|=\frac{\gamma\Delta_0k}{\varepsilon}\sqrt{\frac{\hbar (n_{k}+1)}{2\rho V\omega_{k}}},
\end{equation}
while the matrix element for the absorption process writes $|Q_k^{abs}|=|Q_k^{em}|\sqrt{n_{k}/(n_{k}+1)}$. Using thermal average over the squared matrix element and through Fermi's Golden Rule, one then obtains the relaxation rate of a single TLS to the phonon bath:
\begin{equation}
\label{relax_TLS_ph_generic}
\tau_{ph}^{-1}=\frac{2\pi}{\hbar}\sum_{k}\left[\left\langle\left|Q_{k}^{em}(\Delta_0,\varepsilon)\right|^2\right\rangle_T\delta(\varepsilon-\hbar\omega_{k})+\left\langle\left|Q_{k}^{abs}(\Delta_0,\varepsilon)\right|^2\right\rangle_T\delta(\varepsilon+\hbar\omega_{k})\right].
\end{equation}
Going to the continuum limit, and using the Debye approximation for the phonon modes $\omega_{k}=c_lk$, ($c_l=\sqrt{E/\rho}$ is the longitudinal sound speed) the relaxation rate reads:
\begin{equation}
\label{relax_TLS_ph_continuum}
\tau_{ph}^{-1}=\frac{\gamma^2}{\hbar^2\rho ew c_l^3}\frac{\Delta_0^2}{\varepsilon}\coth\left(\frac{\varepsilon}{2k_BT}\right).
\end{equation}
Note that this result was derived independently in Ref. \cite{Hauer2018}. It may be instructive to write $\tau_{ph}^{-1}$ in reduced variables:
\begin{equation}
\label{relax_TLS_ph_reduced}
\tau_{ph}^{-1}=\frac{\gamma^2}{\hbar^2\rho ew c_l^3}(1-v^2)uk_BT\coth\left(\frac{u}{2}\right),
\end{equation}
where we see that the $u$ dependence outside the coth term must be the same as the one in $T$ because of the transformation $u=\varepsilon/k_BT$. This temperature dependence directly reflects the mechanical damping temperature dependence at very low temperature. We use this property to disregard the contribution of flexural modes: for these, the decomposition of the strain field writes:
\begin{equation}
\label{strain_operator_flexure}
\hat{\mathcal{E}}_f=\sum_{n}\sqrt{\frac{\hbar}{2\rho ew\int\Psi^2_n(z)\mathrm{d}z\omega_{n}}}\left(\hat{a}_{n}^{\dagger}+\hat{a}_{n}\right)y\frac{\partial^2\Psi_n(z)}{\partial z^2}.
\end{equation}
In the string limit, the mode shapes write $\Psi_n=\cos[(n+1)\pi z/l]$ and frequencies are linked to wave numbers $k_n=n\pi/l$ through a linear dispersion relation $\omega_n=c_fk_n$, with $c_f=\sqrt{\sigma_0/\rho}$ the sound speed for the two flexural families (in the string limit they have the same sound speed and mode profiles). Thus $\int\Psi_n^2=l/2$ for all modes, and doing the golden rule calculation for the relaxation of a TLS located at coordinates $(x_t,z)$ in the plan parallel to flexural motion leads to a spatially dependent relaxation rate to the flexural modes:
\begin{equation}
\label{relax_TLS_flexure}
\tau_{ph,f}^{-1}(\varepsilon,\Delta_0,x_t,z)=\frac{2\pi\gamma^2\Delta_0^2x_t^2}{\rho ewlc_f\varepsilon^2}\sum_{n}k_n^3\cos^2(k_nz)\coth\left(\frac{\hbar c_fk_n}{2k_BT}\right)\delta(\varepsilon-\hbar c_fk_{n}).
\end{equation}
The spatial average of this rate, which has to be done in order to obtain the damping rate in the low temperature limit [Eq. (\ref{Damping_lowT_functional_1})], removes the $\cos^2$ term. Besides, since the fundamental flexural mode is $\sim 17.5$ MHz and all modes are equally spaced, at worst (down to 10 mK) roughly 10 modes are thermally activated, which justifies a continuum phonon distribution approximation for simplification. Going to the continuum limit $\sum\rightarrow \int\mathrm{d}k l/\pi$, one obtains:
\begin{equation}
\label{avg_relax_TLS_flexure}
\langle\tau_{ph,f}^{-1}(\varepsilon,\Delta_0)\rangle\propto\Delta_0^2\varepsilon\coth\left(\frac{\varepsilon}{2k_BT}\right),
\end{equation}
which, written in reduced variables, yields a $T^3$ dependence of the damping, essentially negligible at low temperature compared to the $\propto T$ damping due to TLS relaxation to longitudinal modes.
\section{Phonon-driven TLS relaxation in the goalpost cantilever case}
Although we are not aware of a fully comprehensive theory that could explain from first principles the $T^{3/2}$ dependence of the damping rate measured in Ref. \cite{lulla_evidence_2013}, this dependence can be obtained by making some assumptions of a functional form of the phonon-driven TLS relaxation rate. Following Eq. (\ref{TLS_damping_generic}), we write in the low temperature limit, where $\omega\tau_{ph}\gg 1$:
\begin{equation}
\label{Damping_lowT_functional}
\Gamma=C'\int_0^{\infty}\mathrm{d}u\int_0^1\mathrm{d}v\frac{v^2}{1-v^2}\mathrm{sech}^2\left(\frac{u}{2}\right)\tau_{ph}^{-1}(u,v,T),
\end{equation}
where $C'=\iiint_V P_0\gamma^2\mathcal{E}_0^2/2$ is denoted so to emphasize that it differs quantitatively from the $C$ derived earlier, due to different strain fields associated with goalpost cantilever modes. For the same reason, we cannot explicit the phonon driven relaxation rate for lack of a consistent hypothesis about the phonon modes involved in the TLS relaxation process. Nevertheless, from Eq. (\ref{Damping_lowT_functional}) we notice that once $\tau_{ph}$ is written in reduced variables $(u,v)$, its dependence in $T$ is directly reflected by that of the NEMS damping rate. Therefore, based on the damping rate measurements of the goalpost cantilever \cite{lulla_evidence_2013}, we infer a dependence $\tau_{ph}^{-1}(u,v,T)\propto T^{3/2}$. This implies, from the variable transformation $u=\varepsilon/k_BT$ (see ending discussion of the previous paragraph), that a factor $u^{3/2}$ appears, since any temperature dependence outside the $\coth$ term must be canceled when written in explicit variables. In addition, phonon emission and absorption rates, which are summed to obtain the total relaxation rate, must obey detailed balance, i.e. $\tau_{ph,em}^{-1}/\tau_{ph,abs}^{-1}=e^{\varepsilon/k_BT}$. This leads to writing the total relaxation rate $\tau_{ph}^{-1}(u,v,T)=f(v)u^{3/2}\coth\left(\frac{u}{2}\right) T^{3/2}$. Finally, the action of the interaction Hamiltonian (\ref{TLS_ph_pert}) on the TLS states Hilbert space is independent from the phonon bath characteristics and thus is unaffected by the thermal average. The matrix element $Q^{em,abs}$ derived in the previous section for the string case is thus still $\propto \Delta_0/\varepsilon=\sqrt{1-v^2}$. To sum up, one can write the phonon-driven TLS relaxation rate in a semi-phenomenological way:
\begin{equation}
\label{taurel_goalpost}
\tau_{ph}^{-1}=\alpha(1-v^2)u^{3/2}\coth\left(\frac{u}{2}\right)T^{3/2},
\end{equation}
where $\alpha$ is a parameter that contains the TLS-strain coupling energy and ultimately depends on the NEMS geometry and the thermal strain field spatial dependence. Combining the obtained phonon-driven TLS relaxation rate (\ref{taurel_goalpost}) with Eq. (\ref{Damping_lowT_functional}) and performing the integration, one can explicit further the NEMS damping rate:
\begin{equation}
\label{NEMS_damping_cantilever_1}
\Gamma=(1-\sqrt{2}/8)\sqrt{\pi}\zeta\left(\frac{5}{2}\right)\alpha C'T^{3/2}\approx 1.96\,\alpha C' T^{3/2},
\end{equation}
from which $\alpha$ and thus $\tau_{ph}$ may be numerically obtained, using the value of $C'=1.4\times 10^4$ deduced from the damping plateau at high temperatures and leaving $\alpha$ as a fit parameter to match the low temperature $T^{3/2}$ asymptotic law of the damping data in the superconducting state (here we obtain $\alpha\approx 10^7\,\mathrm{K}^{-3/2}.s^{-1}$). These values can then be used when fitting the data in the normal state to our model including electron-driven TLS relaxation (see main text and section below).
\section{TLS-electron interaction}
An incoming conduction electron with wave vector $\textbf{k}$ may be scattered due to the atomic potential to which the TLS contributes. In addition, it may also trigger TLS tunneling (electron-assisted tunneling) by modifying the TLS double-well potential barrier. The Hamiltonian of the TLS-electron interaction writes in the electron Fock state $\{\bigotimes_{k,\eta}|n_{k,\eta}\rangle\}$ $\otimes$ TLS position state basis $\{|\Psi_L\rangle,|\Psi_R\rangle\}$ \cite{coppersmith_low-temperature_1993}:
\begin{equation}
	\label{Hamiltonian_TLS_el_pos}
\hat{\mathcal{H}}_{int,el}^{(p)}=\sum_{k,q,\eta}\left(V_{k,q}^z\hat{c}^{\dagger}_{q,\eta}\hat{c}_{k,\eta}\hat{\sigma}_z+V_{k,q}^x\hat{c}^{\dagger}_{q,\eta}\hat{c}_{k,\eta}\hat{\sigma}_x\right),
\end{equation}
where $\hat{\sigma}_{x,z}$ are the usual Pauli matrices, $\hat{c}^{(\dagger)}_{k,\eta}$ is the fermionic annihilation (creation) operator for an electron of wave vector $k$ and spin $\eta$, and $V_{k,q}^{x,z}$ are longitudinal (z) and transverse (x) interaction constants. One can show that the electron-assisted tunneling interaction parameter satisfy $V_{k,q}^x\sim\langle\Psi_L|\Psi_R\rangle V_{k,q}^z\ll V_{k,q}^z$ and can therefore be neglected.

Let us consider a possible microscopic mechanism of TLS relaxation to the electron bath: for an excited TLS of energy splitting $\varepsilon$ to relax to its ground state, tunneling must occur, which leads to a re-arranging the local atomic configuration and thus to a modification of the local Coulomb potential. An incoming electron with wavevector $\textbf{k}$ will be sensitive to such a potential landscape change within a radius given by the typical screening length $\lambda$ of the medium. This leads to an excitation transfer between the TLS and the electron. The Hamiltonian (\ref{Hamiltonian_TLS_el_pos}) effectively captures this scenario: rewritten in the TLS energy basis,
\begin{equation}
	\label{Hamiltonian_TLS_el_en}
\hat{\mathcal{H}}_{int,el}^{(\varepsilon)}=-\sum_{k,q,\eta}\frac{V_{k,q}^z}{\varepsilon}\hat{c}^{\dagger}_{q,\eta}\hat{c}_{k,\eta}\left(\Delta\hat{\sigma}_z-\Delta_0\hat{\sigma}_x\right),
\end{equation}

which makes TLS transitions appear when electrons are (inelastically) scattered. The transition amplitude to a state of wavevector $q$ associated to e.g. an absorption event and involving an incoming electron of wave vector $k$ writes, assuming state $|q,\eta\rangle$ is intially unoccupied:
\begin{equation}
	\label{matrix_element}
M_{k,q,\eta}(\varepsilon,\Delta_0)\equiv\big\langle n_{q,\eta}+1,n_{k,\eta}-1,e\big|\hat{\mathcal{H}}_{int,el}^{(\varepsilon)}\big|n_{q,\eta},n_{k,\eta},g\big\rangle=\frac{V_{k,q}^z\Delta_0}{\varepsilon}\sqrt{n_{k,\eta}(n_{q,\eta}+1)}
\end{equation}
Using Fermi's Golden Rule, taking into account both absorption and emission, we thus obtain the relaxation rate of a TLS with energy splitting $\varepsilon$ to the electron ensemble:
\begin{equation}
	\label{relax_TLS_el_generic}
\tau^{-1}_{el}(\varepsilon,\Delta_0)=\frac{2\pi}{\hbar}\sum_{k,q,\eta}\left\langle|M_{k,q,\eta}(\varepsilon,\Delta_0)|^2\right\rangle_T\big[\delta(E_k-E_q+\varepsilon)+\delta(E_k-E_q-\varepsilon)\big],
\end{equation}
where $\langle...\rangle_T$ denotes thermal average over the electron reservoir at thermal equilibrium. Applied to the matrix element (\ref{matrix_element}) squared, it makes the product $f(E_k)[1-f(E_q)]$ appear, where $f(E)=1/[\exp([E-\mu]/k_BT)+1]$ is the Fermi-Dirac distribution of the electron bath at equilibrium with temperature $T$ and chemical potential $\mu$. Going to the continuum limit and introducing the (single-spin) electronic density of states (DoS) $n$, one obtains:
\begin{equation}
	\label{relax_TLS_el_continuum}
\tau^{-1}_{el}(\varepsilon,\Delta_0)=\frac{4\pi\Delta_0^2\Omega^2}{\hbar\varepsilon^2}\int_0^{\infty}n(E)f(E)\left(n(E+\varepsilon)[1-f(E+\varepsilon)]V_{E,E+\varepsilon}^2+n(E-\varepsilon)[1-f(E-\varepsilon)]V_{E,E-\varepsilon}^2\right)\mathrm{d}E,
\end{equation}
where $\Omega\sim\lambda^3$ is the effective interaction volume, $\lambda\lesssim 1$ nm being the metallic screening length. Usual approximations allow to simplify this result: due to fermionic occupation rules (appearing through the products $f(E)[1-f(E\pm\varepsilon)]$), only quasiparticle excitations within a bandwidth $k_BT\ll\mu$ around the Fermi level effectively interact with the TLS, which is also a low-energy excitation such that $\varepsilon\ll\mu$. Therefore we can make the approximations $n(E)\approx n_0$ ($n_0\approx 1.07\times 10^{47}\,\mathrm{J}^{-1}\cdot\mathrm{m}^{-3}$ is the single-spin DoS at Fermi level), $V_{E,E^{\prime}}\approx V$. We are left with integrals $\int \mathrm{d}Ef(E)[1-f(E\pm\varepsilon)]$ which are analytical. In the limit $k_BT\ll\mu$,
we finally obtain:
\begin{equation}
	\label{relax_TLS_el_continuum_final}
\tau^{-1}_{el}(\varepsilon,\Delta_0)=\frac{4\pi\Delta_0^2K}{\hbar\varepsilon}\coth\left(\frac{\varepsilon}{2k_BT}\right).
\end{equation}

\section{Crossover temperatures}

To evaluate the crossover temperatures $T_{el}^*$ and $T_{ph}^*$, we equate the damping rate expressions for slow and fast TLS (all TLS satisfying respectively $\omega\tau\gg 1$ and $\omega\tau\ll 1$), for each ensemble of TLS ("neutral" and "charged" ones, see main text). This yields the following crossover temperatures for TLS that relax only to phonons:
\begin{equation}
\label{T_ph}
T_{ph}^*=\frac{3\hbar^2\omega\rho ewc_l^3}{(1-x)\gamma^2\pi k_B},
\end{equation}
where here we take into account the fact that above this temperature, all TLS are fast, and below, only the TLS purely relaxing to phonons are slow, hence the fraction $1-x$ appearing. The obtained crossover temperatures is about 1 K. Meanwhile, the crossover temperature for the other type of TLS is obtained by comparing the low and high temperature limit of damping only for the fraction $x$ of TLS relaxing to both phonons and electrons, which gives a crossover temperature that is independent of $x$:
 \begin{equation}
\label{T_el}
T_{el}^*\approx\frac{3\hbar\omega}{4\pi^2K k_B},
\end{equation}
where we have neglected the phonon contribution to the linear low temperature damping rate. Indeed, with our estimates of TLS-strain and TLS-electron interaction constants, $\tau_{el}^{-1}\gg\tau_{ph}^{-1}$ (at all temperatures since the two relaxation rates have the same temperature and TLS energy dependences). We obtain a crossover temperature $T_{el}^*\approx 0.9$ mK with our estimate of the TLS-interaction constant, which falls outside our reach in temperature.
\section{Additional data}
Here we present additional measurements of a SiN/Aluminum nanomechanical string \cite{zhou2019} having the same mechanical properties as the beam presented in the main text aside from its length (50 $\mu$m). This nanostring, resonating at $3.79$ MHz, is embedded in a $6$ GHz superconducting niobium microwave LC resonator enabling its dispersive readout. This allows us to make measurements free of magnetomotive loading, but prevents operation above 1 K and quenching of the aluminum layer covering the beam to the normal state, since a strong magnetic field cannot be used in these conditions. Yet, we use the frequency shift and damping data to test the STM prediction for pure phonon-assisted relaxation of TLS.

\begin{figure}[ht]
	\centering
	\includegraphics[width=17.2cm]
	{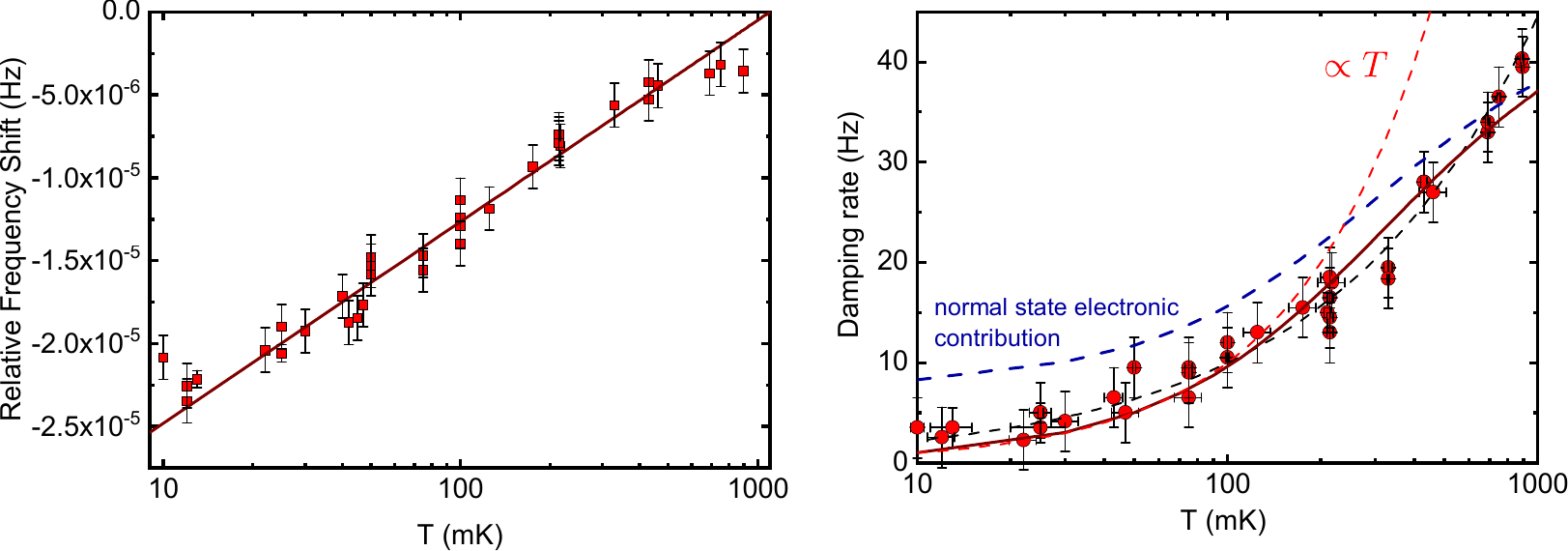}
	\caption{a) Resonance frequency shift of a 50 $\mu$m long SiN nanostring as a function of temperature. The solid line is a logarithmic fit $C\ln(T/T_0)$ b) Damping rate of the NEMS resonator as a function of temperature. The dark red solid line is a fit using Eq. (1) of the main text, with the phonon-driven relaxation rate of TLS derived in this section. The red dashed line is the low temperature asymptotic law, linear in temperature. The blue dashed line represent the extra damping due to conduction electrons if the Aluminum layer were quenched into its normal state. The black dashed line is an empirical $\propto T^{0.65}$ fit.}
	\label{Fig1_SI}
\end{figure}

The resonance frequency shift and damping rate measurements are summed up in Fig. \ref{Fig1_SI}. The data are presented with the back-action contributions (optomechanical spring and damping) \cite{regal08} carefully removed. The frequency shift [Fig. \ref{Fig1_SI}a)] shows a logarithmic dependence in temperature, pointing again towards a TLS mechanism. The extracted $C=5.28\times 10^{-6}$ is low compared to values reported for bulk materials but is reasonably close to values measured in the NEMS literature \cite{hoehne_damping_2010}. The damping rate data are captured with our model using Eq. (\ref{TLS_damping_generic}) and the relaxation rate (\ref{relax_TLS_ph_continuum}). We fit the data to our model [solid dark red curve in Fig. \ref{Fig1_SI}b)] with the same TLS-strain coupling constant $\gamma=9.8$ eV as the one used for the main sample and a TLS density of states $P_0=7.3\times 10^{44}~\mathrm{J}^{-1}\cdot\mathrm{m}^{-3}$, which has the same order of magnitude as the one used for the main sample. The parameters choice is somewhat loose, in the sense that we cannot measure and fit the damping up to a plateau due to an operation limited to very-low temperature. Note that the value of $C$ extracted by extrapolating our data to a high temperature plateau differs from the one measured by frequency shift by only a factor 1.6 (close to that obtained for the main sample), again reflecting possible refinements due to inhomogeneous TLS distribution.

For completeness, we have represented a curve based on a hypothetic electron-assisted damping if one were able to quench the Al layer, using identical interaction strength $K$ and fraction $x$ parameters as those of the main sample. While an excess damping is clearly present below 300 mK, it is not significantly contributing at higher temperatures, being contained in the experimental accuracy. Therefore, a contribution arising from quasiparticles (which will be smaller since their density is smaller than that of normal state electrons) is essentially negligible here, except around 1K, precisely where the purely phononic model falls a little below the measured damping.

\end{document}